\documentclass[3p,times,procedia]{elsarticle}
\usepackage{nupha_ecrc}

\volume{00}

\firstpage{1}

\journalname{Nuclear Physics A}

\runauth{J. Matthew Durham for the PHENIX Collaboration}

\jid{nupha}

\jnltitlelogo{Nuclear Physics A}

\usepackage{amssymb}

\usepackage[figuresright]{rotating}
\usepackage{subcaption}

\begin{document}

\begin{frontmatter}



\dochead{XXVIIth International Conference on Ultrarelativistic Nucleus-Nucleus Collisions\\ (Quark Matter 2018)}

\title{Recent Quarkonia Studies from the PHENIX Experiment}


\author{J. Matthew Durham for the PHENIX Collaboration}

\address{Los Alamos National Laboratory, Los Alamos NM 87545, USA}

\begin{abstract}

Quarkonia suppression in nucleus-nucleus collisions is a powerful tool to probe the density and temperature of the medium created in heavy ion collisions. Forward rapidity measurements in $p(d)$+Au collisions are essential to understand how quarkonia states are affected by initial state effects, formation time, and local particle multiplicity. Earlier measurements in Au+Au collisions showed a stronger suppression of forward $J/\psi$ compared to mid-rapidity results, indicating the possibility of a smaller contribution of regenerated quarkonia states at forward rapidity.  These proceedings report on the latest quarkonia studies performed by the PHENIX collaboration in the rapidity range $1.2<|y|<2.2$.

\end{abstract}


\end{frontmatter}


\section{Introduction}

The suppression of bound heavy quark-antiquark states in collisions of large nuclei has long been expected as a signature of quark-gluon plasma formation \cite{MatsuiSatz}.  However, a variety of other mechanisms responsible for quarkonia suppression that do not require deconfinement have also been proposed, including modifications of the gluon distribution in the nucleus \cite{RamonaOldShadowing}, energy loss of heavy quarks crossing the nuclear remnant \cite{CNM_Eloss}, and late-stage interactions of quarkonia with co-moving hadrons \cite{Capella}.  These effects can be studied experimentally by measurements of charmonia production in small collision systems, where suppression due to plasma formation is not expected to be the dominant effect.

Previous PHENIX results in $d$+Au collisions at the Relativistic Heavy Ion Collider (RHIC) have shown that both open \cite{PPG131,PPG153} and hidden heavy flavor hadrons \cite{PPG109,PPG125} have significant modifications from a simple picture of binary-scaled $p+p$ collisions.  However, there are a variety of models including the above mentioned effects which can successfully reproduce some of these observations \cite{RamonaShadowing,MaCGC,Arleo2013,Ferreiro_comovers}.  Additional experimental data is therefore necessary to further constrain models and identify the sources of these modifications.

\section{New Results in Small Systems} 

\begin{figure}
\centering
\begin{subfigure}{.5\textwidth}
 \centering
  \includegraphics[width=1\linewidth]{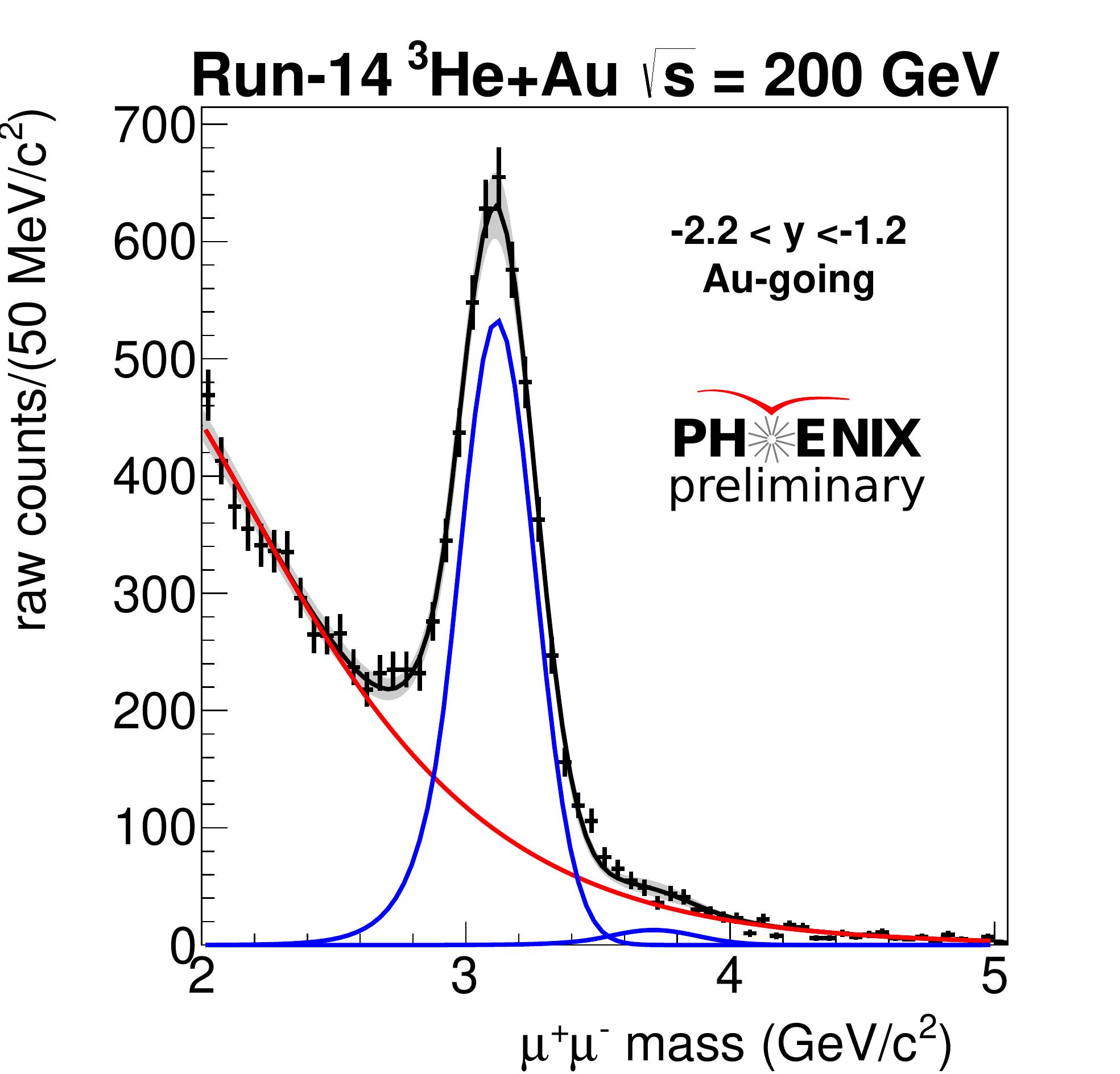}
  \label{fig:sub1}
\end{subfigure}%
\begin{subfigure}{0.5\textwidth}
 \centering
  \includegraphics[width=1\linewidth]{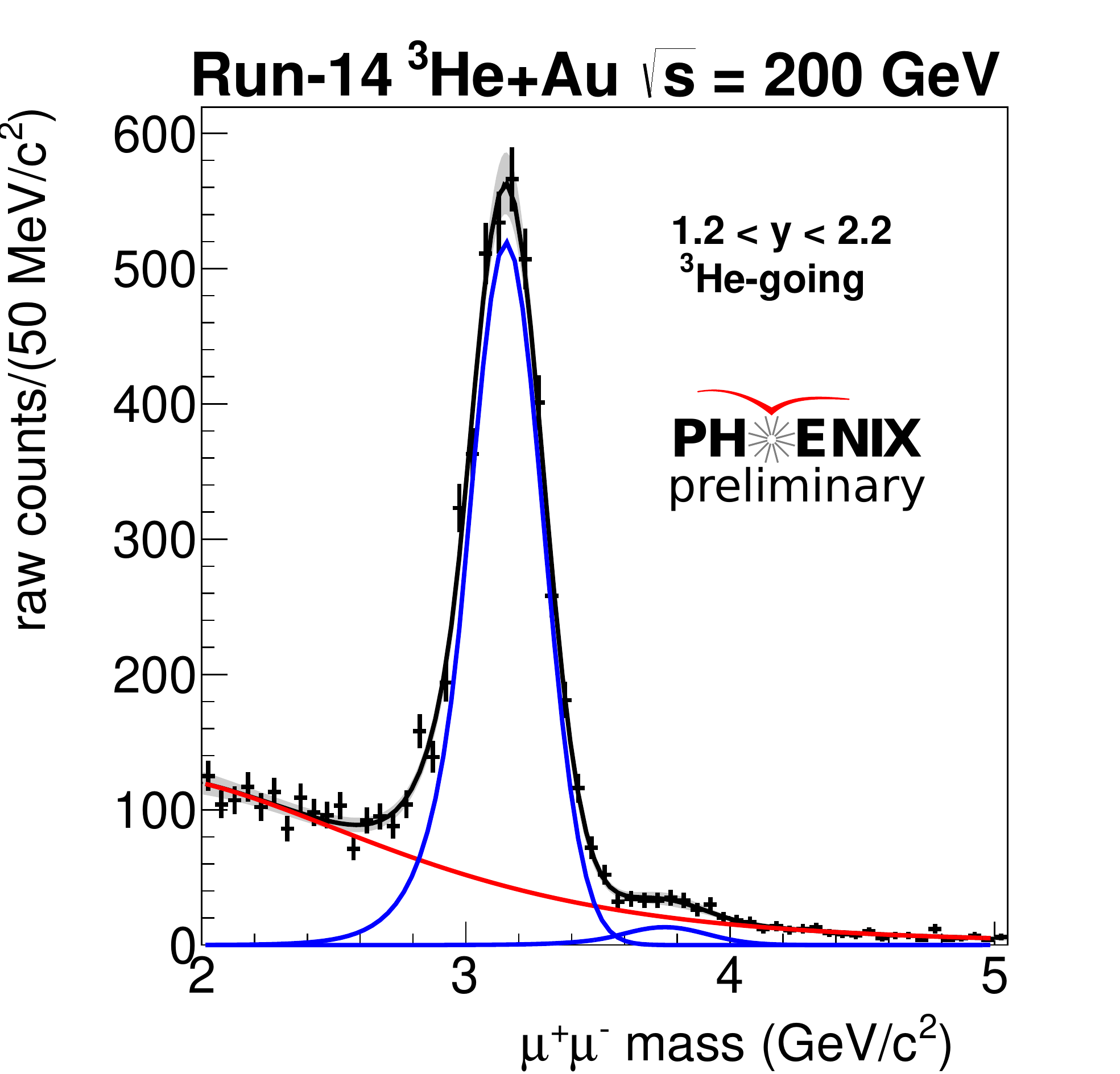}
  \label{fig:sub2}
\end{subfigure}
\caption{Unlike-sign dimuon mass spectra from $^3$He+Au collisions, measured at backward rapidity (left) and forward rapidity (right).  Note the larger background contribution at backwards rapidity, in the Au-going direction.}
\label{fig:fit}
\end{figure}

The flexibility of the RHIC accelerator complex has allowed heavy flavor dynamics across a wide range of collision species to be explored experimentally.  By varying the projectile but keeping the target nucleus the same, the density of co-moving particles can be changed while still sampling the same $x-$range inside the nucleus.  Variation in the nuclear target A samples a different level of shadowing/anti-shadowing in the nuclear gluon distribution.  To this end, the PHENIX collaboration collected data at 200 GeV from $p+$Al and $p+$Au collisions in 2015, and in $^3$He+Au collisions in 2014. In each of these systems, the $J/\psi$ cross section has been measured through decays to dimuons over the forward and backward rapidity intervals $1.2<|y|<2.2$.  The unlike-sign dimuon spectra from $^3$He+Au data, along with the fit used to extract the $J/\psi$ yield, is shown in Fig. \ref{fig:fit}.

\begin{figure}[h]
	\centering
		 \includegraphics[width=0.5\textwidth]{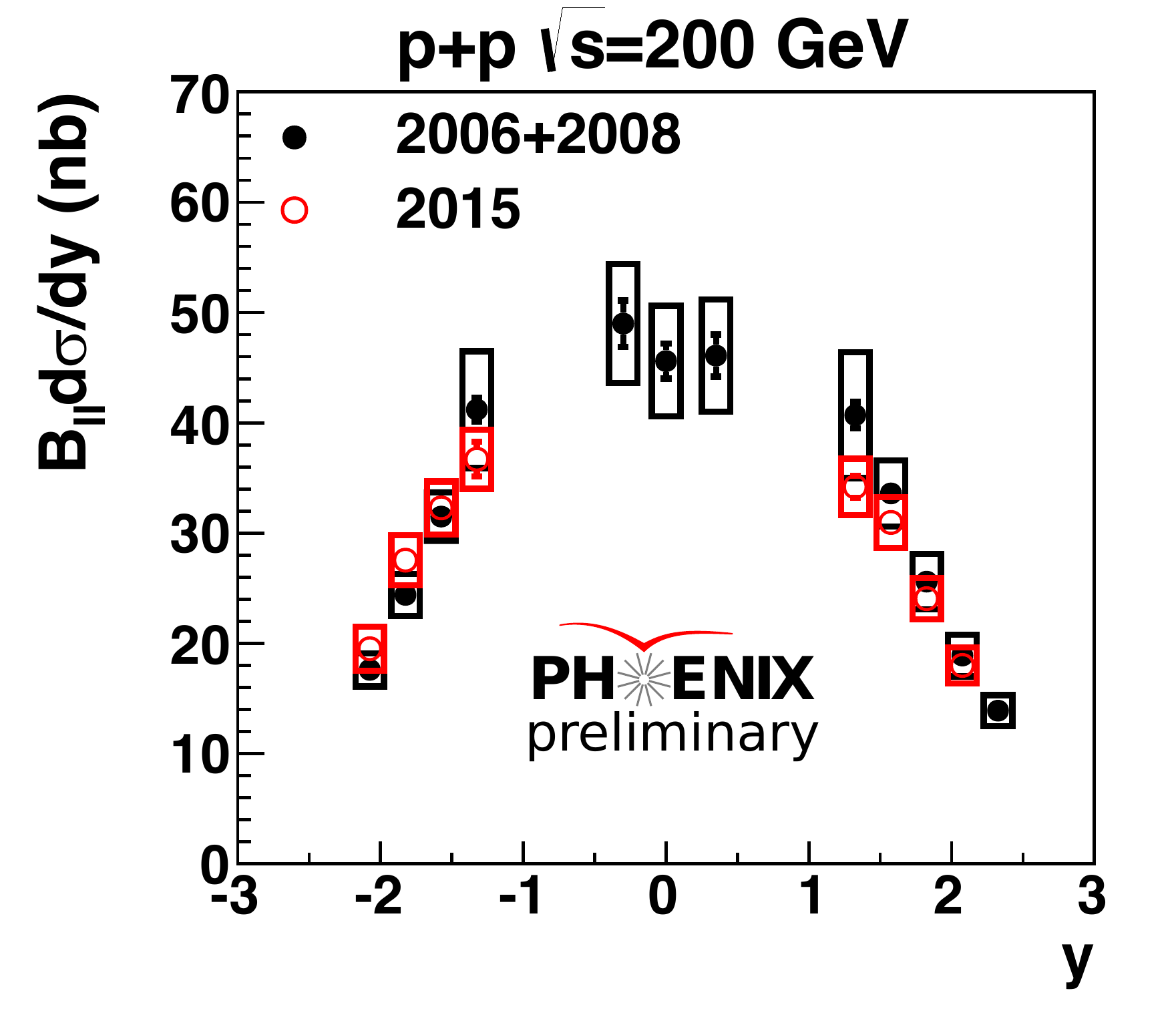}	
	\caption{Comparison of the $J/\psi$ cross section measured in $p+p$ collisions from Run-15 (red) with previous results from \cite{PPG104}.  These datasets, which were recorded with different hadron absorber configurations, show good agreement within uncertainties.}
	\label{fig:pp}
\end{figure}

Data from $p+p$ collisions, which gives the baseline for interpreting nuclear effects, was also recorded in 2015.  This data serves as an important cross check with previous measurements, as additional hadron absorber material in front of the PHENIX muon arms was introduced between these measurements to provide additional background reduction for the $W$ measurement program \cite{PPG120}.  The $J/\psi$ cross section from the 2015 $p+p$ data, and from previous measurements with less absorber material \cite{PPG104} is shown in Fig.\ref{fig:pp}.  We see that these measurements agree within uncertainties.

\begin{figure}[h]
\centering
\begin{subfigure}{.36\textwidth}
  \centering
  \includegraphics[width=1\linewidth]{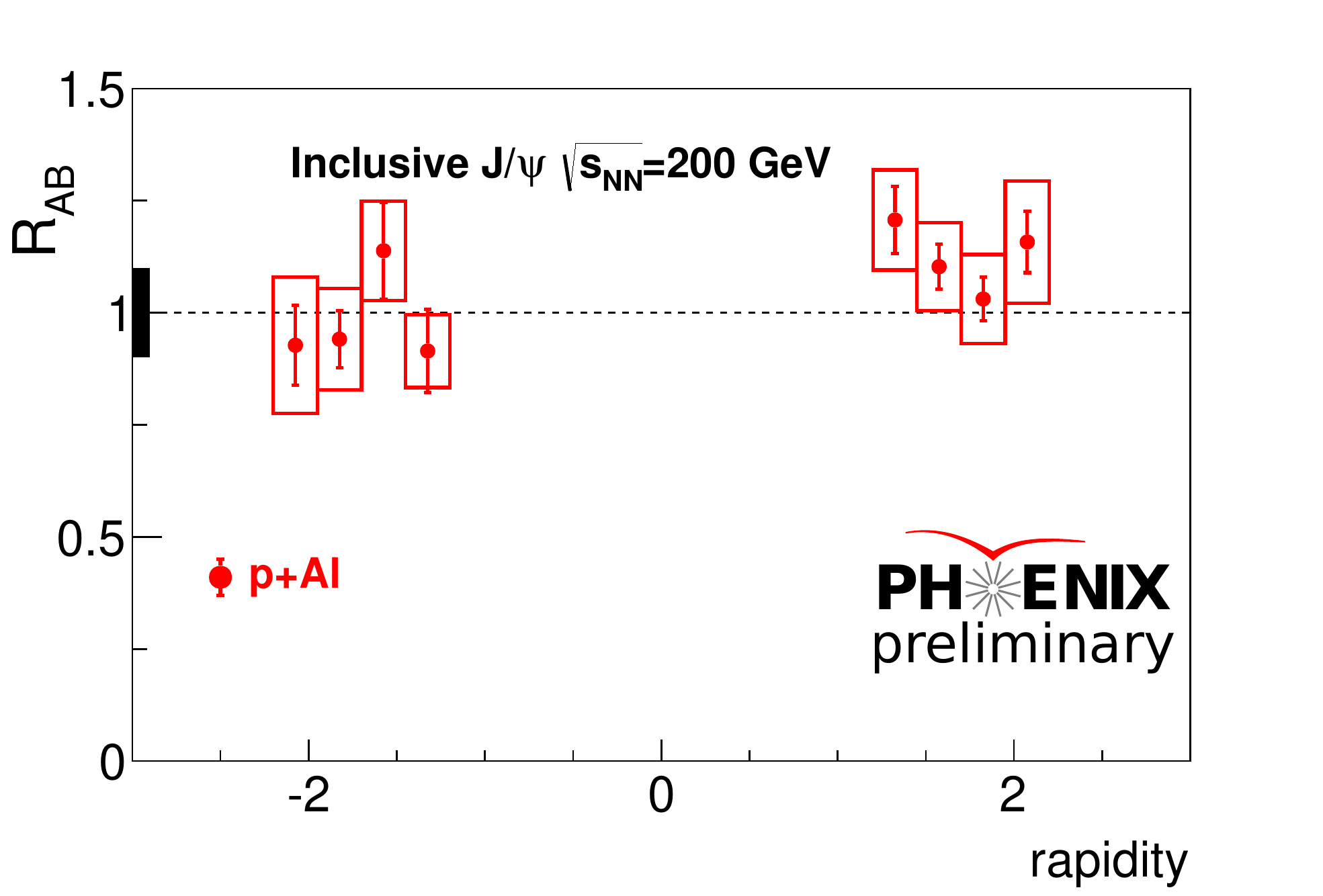}
  \label{fig:sub1}
\end{subfigure}%
\begin{subfigure}{.36\textwidth}
  \centering
  \includegraphics[width=1\linewidth]{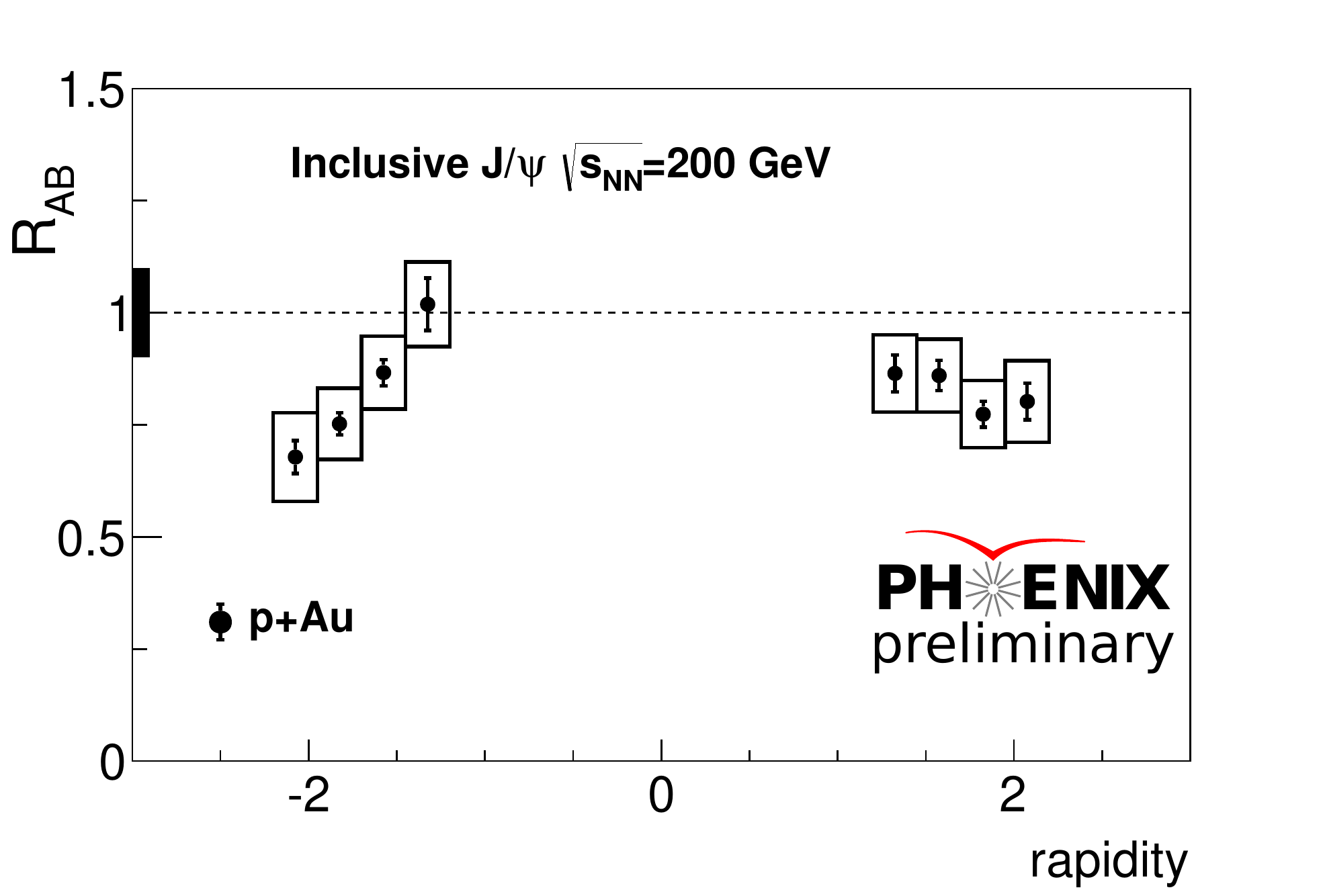}
  \label{fig:sub2}
\end{subfigure}%
\begin{subfigure}{0.36\textwidth}
  \centering
  \includegraphics[width=1\linewidth]{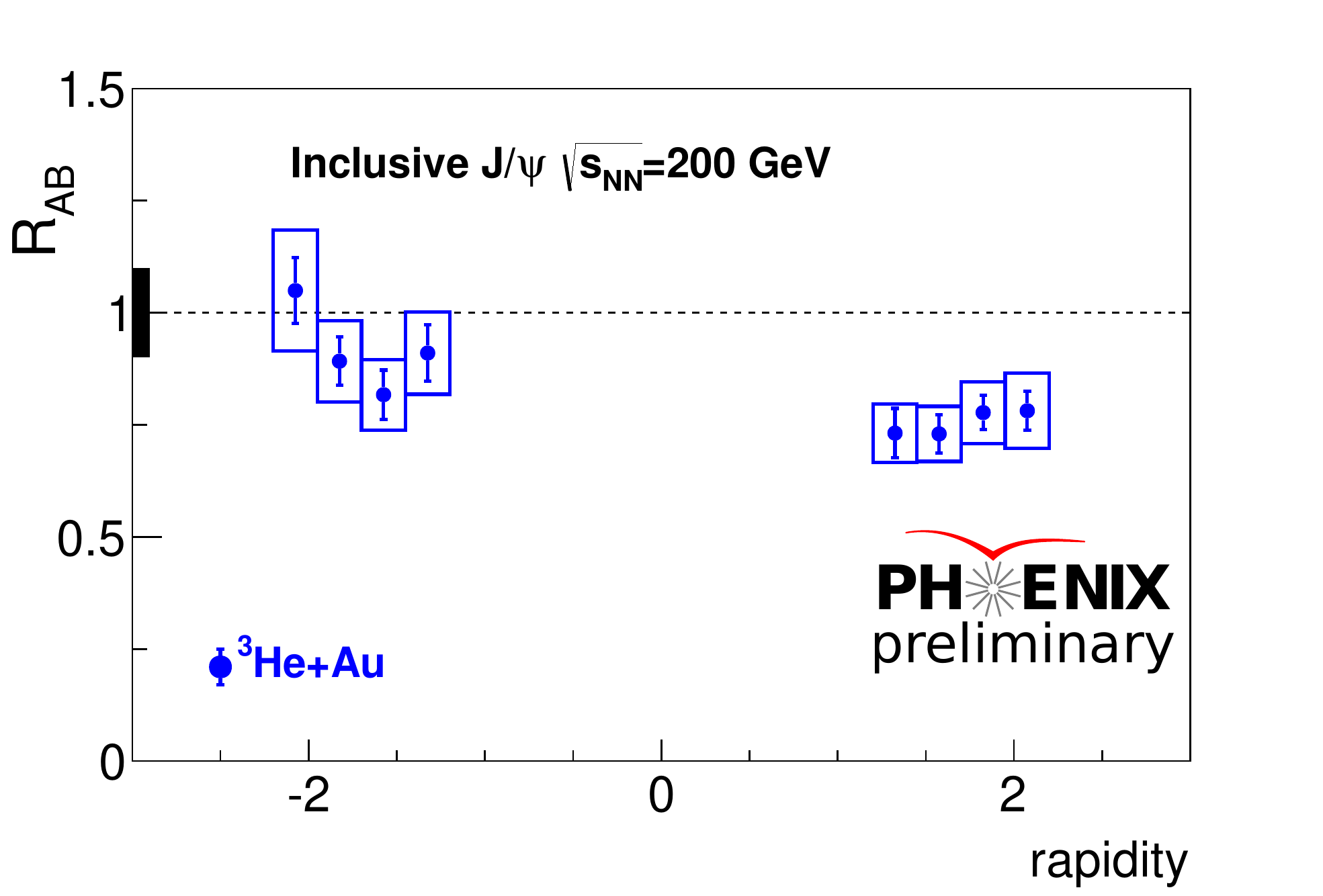}
  \label{fig:sub3}
\end{subfigure}
\caption{The rapidity dependence of the $J/\psi$ nuclear modification factor $R_{AB}$ for $p$+Al, $p$+Au, and $^3$He+Au collisions. }
\label{fig:rap}
\end{figure}

The $J/\psi$ cross section as a function of rapidity was also measured for the $p+$Al, $p+$Au, and $^3$He+Au systems, and is directly compared to the $p+p$ data by computing the nuclear modification factor $R_{AB}$ as shown in Fig. \ref{fig:rap}.  We see that in $p+$Al collisions, the nuclear modification is small, and compatible with unity within uncertainties.  As we move to the larger $p+$Au and $^3$He+Au collision systems, there is an increasing suppression at forward rapidity.  In this rapidity region, $J/\psi$ measured in the PHENIX muon arms sample an $x-$range of  $x\approx$5$\times$10$^{-3}$, which is in the gluon shadowing region.  There is also a relatively low density of co-moving hadrons in the forward region, so late stage breakup from interactions which occur outside the nucleus is expected to be small.  We note that in this region, light hadrons are suppressed by a similar amount \cite{JasonProc}.

At backward rapidity, we see some evidence of some moderate suppression in $p+$Au collisions, and less so in $^3$He+Au collisions.  In this region, the $x$-range being sampled by charmonia production is in the gluon anti-shadowing region, and suppression due to parton distribution function modification is not expected.  Since this is the A-going direction, the amount of co-moving hadrons is significantly higher in this rapidity region, and late stage interactions that may be present would therefore have a larger magnitude.  In contrast to the $J/\psi$, charged hadrons in this region display a large enhancement at moderate $p_{T}$, by a factor of $\sim$1.5 in 0-100$\%$ $p+$Au collisions.  Light hadrons are not susceptible to breakup so the fact that we observe different behavior here may indicate that this additional suppression mechanism which only effects heavy quarkonia is dominant.  However, given the differences in light and heavy quark production mechanisms, additional guidance from theory is needed to fully understand the various contributions.


\section{Comparison with Large Systems}

\begin{figure}
\centering
\begin{subfigure}{.55\textwidth}
  \centering
  \includegraphics[width=1\linewidth]{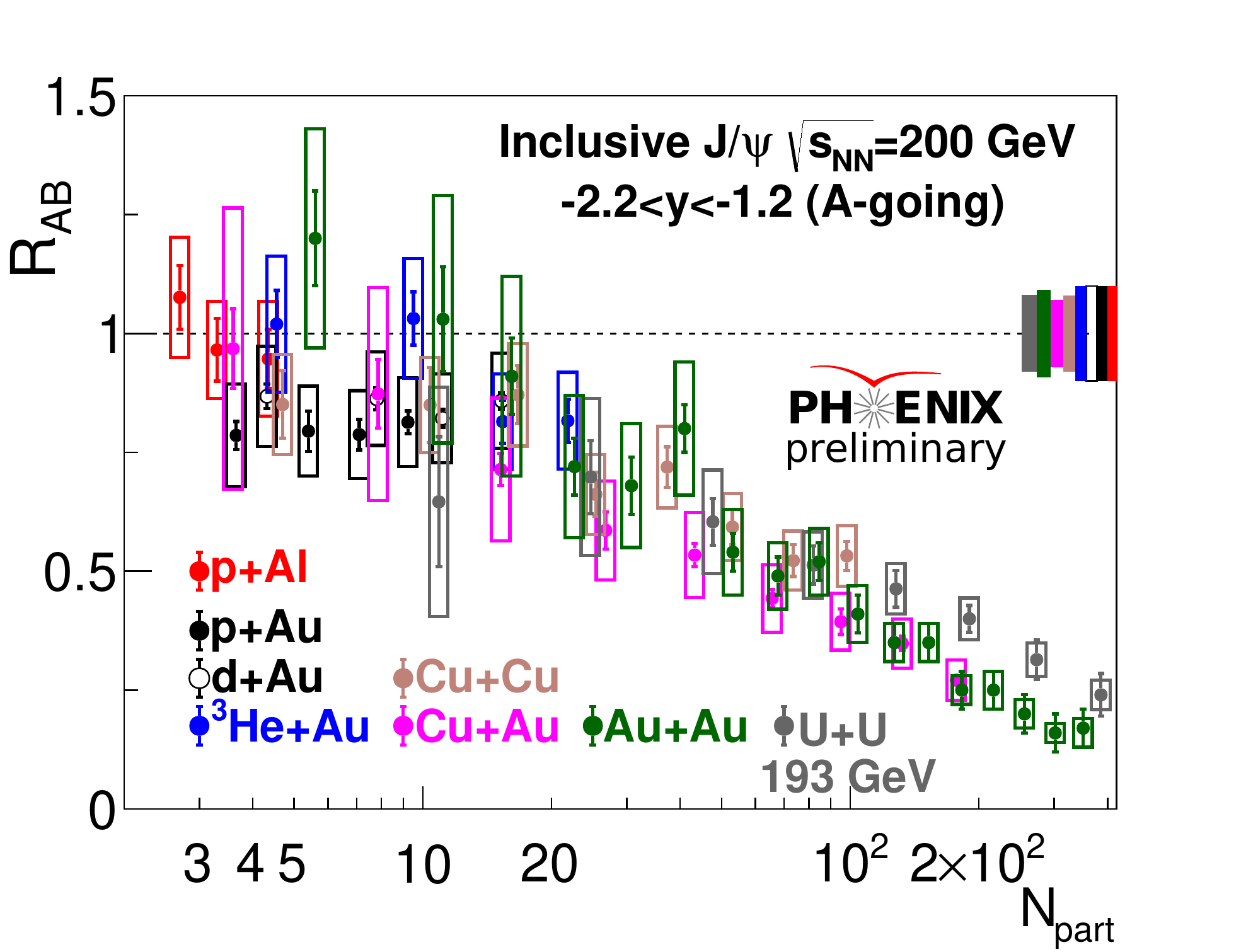}
  \label{fig:sub1}
\end{subfigure}%
\begin{subfigure}{0.55\textwidth}
  \centering
  \includegraphics[width=1\linewidth]{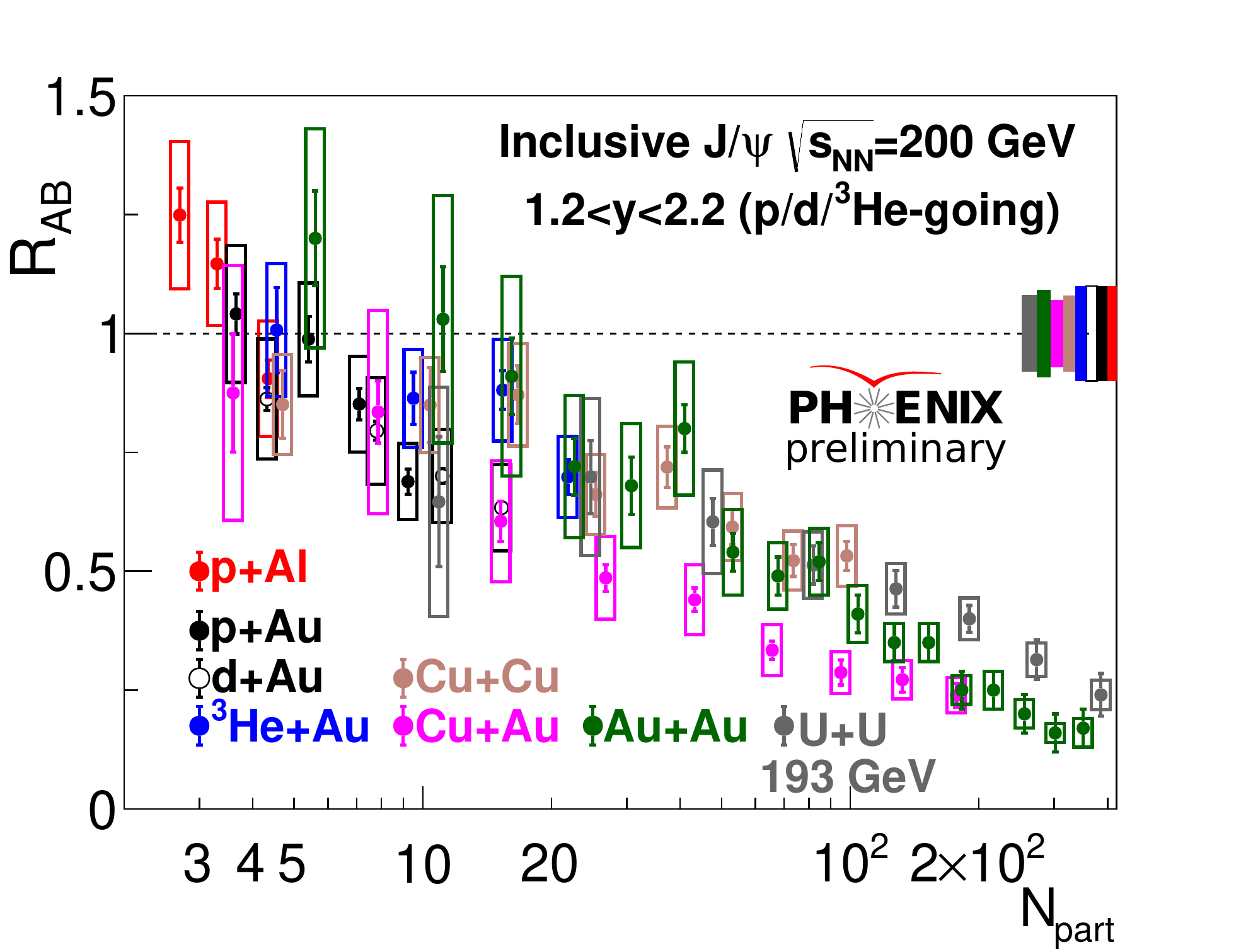}
  \label{fig:sub2}
\end{subfigure}
\caption{PHENIX data on the $J/\psi$ nuclear modification factor as a function of the number of participating nucleons, at backward rapidity (left) and forward rapidity (right). }
\label{fig:all}
\end{figure}

In addition to these small systems, the PHENIX experiment has measured the $J/\psi$ cross section in Cu+Cu \cite{PPG071}, Cu+Au \cite{PPG163}, Au+Au \cite{PPG119}, and U+U \cite{PPG172} collisions.  Taken together, these measurements cover $N_{part}$ from 2 to nearly 400, probing a wide range of system size and temperature.  The measurements on these systems at forward and backward rapidity are shown in Fig. \ref{fig:all}.  For symmetric collision systems, the same data is shown in each panel.  Across all collision systems, there is similar suppression at similar values of $N_{part}$.  In both rapidity regions, we see a suppression develop as $N_{part}$ increases, indicating that additional suppression effects are coming into play as the reaction volume increases.

\section{Summary}

The PHENIX Experiment has measured $J/\psi$ production across a wide range of collision systems, where the effects of competing suppression mechanisms are expected to have different strengths.  Further exploration of these data sets, including measurements of the $p_{T}$ dependence of the $J/\psi$ nuclear modification, is underway.  While PHENIX is no longer recording data, there are large data sets on tape from the Run-14 and Run-16 Au+Au that are currently undergoing analysis.  The high statistics measurements from this data will provide additional constraints on suppression mechanisms in large collision systems.





\bibliographystyle{elsarticle-num}
\bibliography{PPG188}







\end{document}